\documentstyle[twoside,fleqn,espcrc2,epsfig]{article}

\newcommand{\AmS}{{\protect\the\textfont2
  A\kern-.1667em\lower.5ex\hbox{M}\kern-.125emS}}

\title{New Estimates On Various Critical/Universal Quantities \\
       of the 3D Ising Model
        \thanks{Poster presented by K. Pinn, preprints 
        HUB-EP-97/52, MS-TPI-97-10}}
\author{M. Hasenbusch\address{Fachbereich Physik, 
        Humboldt-Universit\"at zu Berlin, \\ 
        Invalidenstr.~110, D-10099 Berlin, Germany}
        and 
        K. Pinn\address{Institut f\"ur Theoretische Physik I,
        Universit\"at M\"unster, \\
        Wilhelm-Klemm-Str.~9, D-48149 M\"unster, Germany}}
       
\begin{document}

\begin{abstract}
We present estimates for the 3D Ising model on the cubic lattice,  both
regarding interface and bulk properties.  We have results for the
interface tension, in particular the amplitude $\sigma_0$ in the
critical law $\sigma = \sigma_0 \, t^{\mu}$, and for the universal
combination $R_{-} = \sigma \, \xi^2$. Concerning the bulk properties,
we estimate   the specific heat universal amplitude ratio $A_+/A_-$,
together with the exponent $\alpha$,  the nonsingular background of
energy and specific heat at criticality,  together with the exponent
$\nu$.  There are also results for the universal combination $f_{\rm s} \,
\xi^3$, where $f_{\rm s}$ is the singular part of the free energy. 
Details can be found in \cite{isiface} (interface) and  \cite{bulk} (bulk). 
\end{abstract}

\maketitle
\section{INTRODUCTION}

In this conference contribution,  we summarize recent
efforts~\cite{isiface,bulk} to determine  various critical quantities of
the 3D Ising universality class using the Monte Carlo (MC) method. These
efforts are worthwhile for at least three reasons: 
1) There are many
systems in nature belonging to the Ising universality class. 
Thus a comparison of
theoretical predictions with experimental results is possible.  2)
Getting precise estimates is still a challenge, both from the
algorithmic side and from a theoretical point of view. MC data with
increasing statistical precision  put standard assumptions to test and
may  require more and more refined theoretical modelling. 3) MC
precision has now become competitive with, e.g., high/low temperature
series expansions, the $\epsilon$-expansion and 3D field theoretic
calculations. Comparing the results is interesting and might point out
sources of systematic errors in either method.

\section{INTERFACE PROPERTIES}

Properties of interfaces separating extended domains of 
different phases are of great interest, both from the
experimental and the theoretical point of view.
For recent experimental work measuring observables that are 
related to and can be compared with quantities studied in 
the present work, see \cite{mainzer,isiface}.

\vskip5mm
\noindent
Figure 1 \\
Lattice geometry
\begin{center}
{~}
\epsfig{file=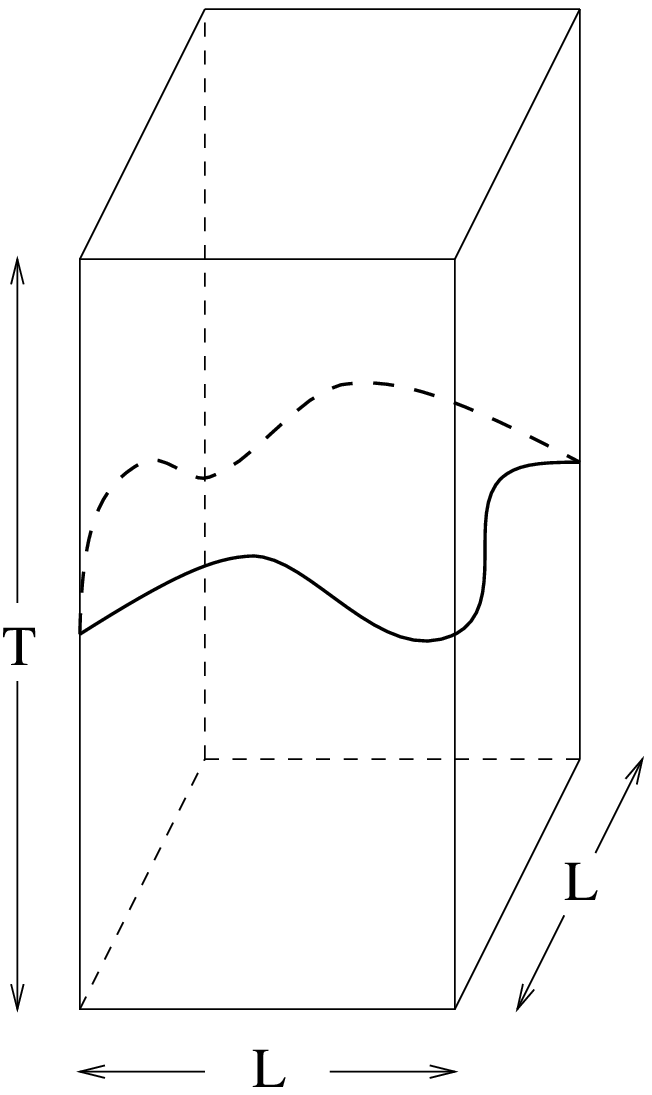,height=4.0cm}
\end{center}

\vskip2mm

We consider the 3D Ising model on a cubic lattice 
of size $L \times L \times T$
with spins $s_x= \pm 1$, 
Hamiltonian $H(s)=-\beta\, \sum_{<x,y>} s_x s_y$, where
$<\!xy\!>$ denotes a pair of nearest neighbour sites on the lattice,
and partition function $Z= \sum_s \exp[-H(s)]$.
By $Z_p$ we denote the partition function with periodic
boundary conditions in all three directions. 
Antiperiodic boundary
conditions in $t$-direction favour the occurence of 
interfaces (perpendicular to the $t$-axis), see figure~1.
The corresponding partition function is denoted by $Z_a$. 
If we can assume that there is exactly one interface in the
system, its free energy is given by 
\begin{equation}
F_s = F_a - F_p + \ln T \, ,
\end{equation}
where $\ln T$ takes care of the free motion of the interface
in the $t$-direction.

We obtained the interface free energies by integration 
of the interface energy 
$E_s = \langle H \rangle_a - \langle H \rangle_p$: 
\begin{equation}
F_s = F_s(\beta_o) + \int_{\beta_o}^{\beta} 
d\beta' \, E_s(\beta') \, .
\end{equation}
Various state of the art algorithms were employed:
the boundary flip cluster algorithm for the determination 
of $F_s(\beta_0)$, the Wolff single cluster, the
Ising interface cluster and the multispin coded demon
algorithm for the determination of $E_s$. Detailed descriptions
and the appropriate references can be found in~\cite{isiface}.

Interface tensions $\sigma$ were then obtained by fitting the free 
energies with either 
\begin{equation}\label{simple}
F_s \simeq C_s + \sigma \, L^2  
\end{equation} 
or~\cite{twoloop}
\begin{equation}\label{two-loop}
F_s \simeq C_s + \sigma \, L^2 \, + 1/(4\sigma L^2) \, .
\end{equation}
We chose lattice sizes $32^3$, $48^3$, $64^3$ and 
$96^3$. Data were taken on a dense grid of $\beta$ values
in the range $0.223 \dots 0.23$. 
In table \ref{tab1} we quote a few of our fit results
for the interface tension.
\begin{table}[hbt]
\setlength{\tabcolsep}{0.95pc}
\newlength{\digitwidth} \settowidth{\digitwidth}{\rm 0}
\catcode`?=\active \def?{\kern\digitwidth}
\caption{Selection of $\sigma$ estimates}
\label{tab1}
\begin{tabular}{lll}
\hline
fit             &
$\beta=0.2240$   &
$\beta=0.2255$ \\
\hline
fit2    &  0.004784(9)  & 0.008810(9)   \\
96vs64  &  0.004799(17) & 0.008822(18)  \\
fit3    &  0.004782(6)  & 0.008801(6)   \\
\hline
\end{tabular}
\end{table}
Here, ``fit2'' labels a fit with eq.~(\ref{simple}), discarding 
the $L=32$ data, under the name of ``96vs64'' goes a determination
of $C_s$ and $\sigma$ from the $L=64$ and $L=96$ data alone, 
and ``fit3'' refers to a fit of data from all four lattice sizes 
with eq.~(\ref{two-loop}). The results from the different
fit types are fairly consistent.

Our next task was to fit the interface tension results
to the critical law 
\begin{equation}
\sigma(\beta) \simeq \sigma_0 \, t^{\mu} (1 + a_{\theta} t^{\theta}
+ a_1 \, t) \, .
\end{equation}
Here we included the leading corrections to scaling terms.
We compared the results stemming from the two definitions of 
the reduced temperature, 
$t_1= \beta/\beta_c -1$ and 
$t_2= 1 - \beta_c/\beta$. 
Furthermore, we fixed the following parameters:
$\beta_c=0.2216544(6)$ \cite{talapov}, $\mu=1.262$, and 
$\theta=0.51$ \cite{talapov}. 
These were the most precise estimates we could find in
the literature. The fit results are quoted in table~\ref{tab2}, again
for three different fit schemes. 

\begin{table}[hbt]
\setlength{\tabcolsep}{0.3pc}
\caption{Fit results for $\sigma_0$, $a_{\theta}$ and $a_1$}
\label{tab2}
\begin{tabular}{llllr}
\hline
\multicolumn{1}{l}{fit} &
\multicolumn{1}{l}{$t$-def} &
\multicolumn{1}{l}{$\sigma_0$} &
\multicolumn{1}{l}{$a_{\theta}$}  &
\multicolumn{1}{l}{$a_1$} \\
\hline
fit2   & $t_1$ & 1.549(11)  & -0.409(72)  &  0.01(20) \\
             & $t_2$ & 1.549(11)  & -0.397(76)  &  1.13(21) \\
\hline
fit3   & $t_1$ & 1.5428(73) &  -0.376(49) & -0.06(14) \\
             & $t_2$ & 1.5421(73) &  -0.362(51) &  1.06(14) \\
\hline
96vs64 & $t_1$ & 1.571(20) & -0.57(13) &   0.47(36) \\
             & $t_2$ & 1.571(21) & -0.56(13) &   1.61(37) \\
\hline
\end{tabular}
\end{table}
Note that the $\sigma_0$ results
for the two $t-$definitions are always consistent with each other.
$a_1$ should have a jump of $\mu\approx 1.26$ which is obeyed within
the errors.
Taking into account the various sources of uncertainty (including
the variation with the type of fit) we arrive at the final estimate
$\sigma_0 = 1.55(5)$.

We finally employed our $\sigma$-data in an estimation of 
the universal ratio $R_{-}$. Including corrections to scaling, 
it can be defined through
\begin{equation}
R_{-} = \sigma(\beta) \, \xi(\beta)^2 - c \, \xi^{-\omega} \, , 
\end{equation}
where $\beta \rightarrow \beta_c$ from above.
$\omega$ is the correction to scaling exponent. We fixed it to
0.81(5)~\cite{talapov}.
The $\xi$-data were taken from~\cite{martin-michele}. Our result is
$R_{-}= 0.1040(8)$, where the error is mainly due to the uncertainty
in the exponent $\omega$.
A comparison with some results from the  literature is
given in~\cite{isiface}.

\section{BULK PROPERTIES}

We consider a 3D Ising model on a cubic lattice of size $L^3$,
with periodic boundary conditions in all three directions.
Important quantities are the free energy per volume
\begin{equation}
f = -1/(3\,L^3) \ln Z \, ,
\end{equation}
the energy, $E=-df/d\beta$ and the specific heat,
$C= dE/d\beta$. The quantities are split in 
a nonsingular and a singular part, e.g., $E=E_{\rm ns} + E_{\rm s}$. 
By standard arguments the following finite size scaling laws 
are expected to hold:
\begin{eqnarray}\label{atcrit}
E &\simeq& E_{\rm ns} + const_{E} \, L^{-d+1/\nu} \nonumber \\ 
C &\simeq& C_{\rm ns} + const_{C} \, L^{-d+2/\nu} 
\end{eqnarray}
at criticality, and 
\begin{equation}\label{offcrit}
E \simeq E_{\rm ns} - C_{\rm ns} \, \beta_c \, t
\, \mp \,  A_{\pm} \, \beta_c \,
\frac{ |\, t\, |^{1-\alpha}}{1-\alpha}  \, ,
\end{equation}
above and below the transition.
Using state of the art algorithms we obtained estimates for
energy and specific heat at $\beta_c= 0.2216544$ for $L$ ranging
from 12 to 112. The relative error of the energy is of order $10^{-5}$
throughout. 
We then did a combined fit of energy and specific heat to eq.~(\ref{atcrit}).
Our results for $\nu$, $E_{\rm ns}$ and $C_{\rm ns}$ are given 
in table~\ref{tab3}.
\begin{table}[hbt]
\setlength{\tabcolsep}{0.3pc}
\caption{Fit results from critical energy and specific heat}
\label{tab3}
\begin{tabular}{lllll}
\hline
$L_{min}$ & $X$   & $\nu$  & $E_{\rm ns}$  &  $C_{\rm ns}$ \\
\hline
 12  & 5.6  &0.6316(4) &0.330229(5)&-12.13(33) \\
 16  & 2.25 &0.6315(8) &0.330218(7)&-11.83(61) \\
 20  & 0.75 &0.6308(10)&0.330209(8)&-11.12(76)\\
\hline
\end{tabular}
\end{table}
$L_{min}$ is the smallest lattice size used for the fit, and
$X$ denotes $\chi^2$ per degree of freedom.
We find the precision of the $\nu$-estimate of the last line
of the table quite remarkable. 

In the off critical case, we always aimed at the infinite volume
limit of the energy. Our data cover the range from 
$\beta = 0.21971$ to $\beta=0.2230$. Lattice sizes up to $L=128$
were employed.
The results of fits with eq.~(\ref{offcrit}) are given in 
table~\ref{tab4}.
\begin{table}[hbt]
\setlength{\tabcolsep}{0.6pc}
\caption{Fit results from off critical energy}
\label{tab4}
\begin{tabular}{lll}
\hline
                 &  $t_1$      & $t_2$         \\
\hline
{$\alpha$}       & \phantom{-}0.1115(37) &\phantom{-1}0.1047(48) \\
{$A_{+}/A_{-}$}  & \phantom{-}0.550(12)  &\phantom{-1}0.567(16)  \\
{$E_{\rm ns}$}   & \phantom{-}0.33029(3) &\phantom{-1}0.33030(4) \\
{$C_{\rm ns}$}   & -9.86(80)             &-11.37(1.17)  \\
\hline
\end{tabular}
\end{table}
Again we used two definitions of $t$, namely $t_1 = 1-\beta/\beta_c$ 
and $t_2= \beta_c/\beta - 1$. There is a slight mismatch of the 
estimate for $E_{\rm ns}$ with the estimates from the critical
data in table~\ref{tab3}. It is most likely due to corrections
to scaling that have not been taken into account.

We finally estimated the universal constant $f_{\rm s} \, \xi^3$
on both sides of the transition. Estimates for $f_{\rm s}$ were 
obtained by integration over $\beta$ of the singular part
of the energy, obtained from the approximation
$E_s \approx E - E_{\rm ns} - C_{\rm ns} (\beta - \beta_c)$. 
Using our background estimates $E_{\rm ns}$ and $C_{\rm ns}$ 
together with the correlation lengths given in~\cite{martin-michele}, 
we obtained 
$f_{\rm s} \, \xi^3 = 0.0355(15)$ for $\beta \nearrow \beta_c$, 
and
$f_{\rm s} \, \xi^3 = 0.0085(2)$ for $\beta \searrow \beta_c$.

\section{CONCLUSIONS}
For various interface and bulk quantities we could significantly
improve on older numerical estimates. 
Especially interesting is the quite precise determination
of $\nu$ from the finite size scaling of the energy and 
specific heat at criticality.

\end{document}